\documentstyle [12pt] {article}

\newcommand{\gtwid}{\mathrel{\raise.3ex\hbox{$>$\kern-.75em\lower1ex
\hbox{$\sim$}}}}
\newcommand{\ltwid}{\mathrel{\raise.3ex\hbox{$<$\kern-.75em\lower1ex
\hbox{$\sim$}}}}
\newcommand{\beq}{\begin{equation}}
\newcommand{\eeq}{\end{equation}}
\newcommand{\beqs}{\begin{eqnarray}}
\newcommand{\eeqs}{\end{eqnarray}}

\catcode`@=11
\@addtoreset{equation}{section}
\@addtoreset{equation}{subsection}
\def\theequation{\ifnum\value{section}=0 \arabic{equation}\ignorespaces
\else \ifnum\value{section}=-1 A.\arabic{equation}\ignorespaces
\else \ifnum\value{subsection}=0 \thesection.\arabic{equation}\ignorespaces
\else \thesection.\arabic{subsection}.\arabic{equation}\ignorespaces
                           \fi
                      \fi
                 \fi}
\catcode`@=12

\begin{document}

\def\thefootnote{\fnsymbol{footnote}}
\baselineskip 6.0mm

\begin{flushright}
\begin{tabular}{l}
ITP-SB-95-51    \\
December, 1995
\end{tabular}
\end{flushright}

\vspace{8mm}
\begin{center}
{\Large \bf Some New Results on Yang-Lee Zeros }

\vspace{3mm}

{\Large \bf of the Ising Model Partition Function}

\vspace{16mm}

\setcounter{footnote}{0}
Victor Matveev\footnote{email: vmatveev@insti.physics.sunysb.edu}
\setcounter{footnote}{6}
and Robert Shrock\footnote{email: shrock@insti.physics.sunysb.edu}

\vspace{6mm}
Institute for Theoretical Physics  \\
State University of New York       \\
Stony Brook, N. Y. 11794-3840  \\

\vspace{16mm}

{\bf Abstract}
\end{center}

   We prove that for the Ising model on a lattice of dimensionality
$d \ge 2$, the zeros of the partition function $Z$ in the complex $\mu$ plane
(where $\mu=e^{-2\beta H}$) lie on the unit circle $|\mu|=1$ for a wider
range of $K_{n n'}=\beta J_{nn'}$ than the range $K_{n n'} \ge 0$
assumed in the premise of the Yang-Lee circle theorem.  This range includes
complex temperatures, and we show that it is lattice-dependent.
Our results thus complement the Yang-Lee theorem, which
applies for any $d$ and any lattice if $J_{nn'} \ge 0$.
For the case of uniform couplings $K_{nn'}=K$, we show that these zeros lie
on the unit circle $|\mu|=1$ not just for the
Yang-Lee range $0 \le u \le 1$, but also for (i) $-u_{c,sq} \le u \le 0$ on the
square lattice, and (ii) $-u_{c,t} \le u \le 0$ on the triangular lattice,
where $u=z^{1/2}=e^{-4K}$, $u_{c,sq}=3-2^{3/2}$, and $u_{c,t}=1/3$.
For the honeycomb,
$3 \cdot 12^2$, and $4 \cdot 8^2$ lattices we prove an exact symmetry of the
reduced partition functions,
$Z_r(z,-\mu)=Z_r(-z,\mu)$. This proves that the zeros of
$Z$ for these lattices lie on $|\mu|=1$ for $-1 \le z \le 0$ as well as the
Yang-Lee range $0 \le z \le 1$.  Finally, we report some new results on the
patterns of zeros for values of $u$ or $z$ outside these ranges.

\pagestyle{empty}
\newpage

\pagestyle{plain}
\pagenumbering{arabic}
\renewcommand{\thefootnote}{\arabic{footnote}}
\setcounter{footnote}{0}

  The Ising model has long served as a simple prototype of a
statistical mechanical system which (for $d$ greater than the lower critical
dimensionality $d_{\ell.c.d.}=1$) undergoes a second-order phase
transition with associated spontaneous symmetry breaking
and long range magnetic order.  The general zero-field (spin 1/2) Ising model
on a lattice $\Lambda$ at temperature $T$ and external magnetic field $H$ is
defined by the partition function
\beq
Z = \sum_{\{\sigma_n\}} e^{-\beta {\cal H}}
\label{zfun}
\eeq
with the Hamiltonian
\beq
{\cal H} = - \sum_{<nn'>} \sigma_n J_{n n'}\sigma_{n'} - H \sum_n \sigma_n
\label{ham}
\eeq
where $\sigma_n = \pm 1$ are the spin variables on each site $n$ of the
lattice, $J_{n n'}$ is the spin-spin exchange constant,
and the units are defined such that the
magnetic moment which would multiply $H\sum_n \sigma_n$ is unity.  We shall
prove a general result for the case of arbitrary $J_{n n'}$ (connecting any
two sites $n$ and $n'$) and then concentrate on the usual nearest-neighbor
model with uniform coupling $J_{n n'} = J \delta_{n, n' \pm e_j}$,
where $e_j$ is a lattice vector.
We use the standard notation $\beta = (k_BT)^{-1}$,
$K = \beta J$, $h = \beta H$, $z = e^{-2K}$,
$u = z^2 = e^{-4K}$, and $\mu = e^{-2h}$.  The
reduced free energy (per site) is
$f = -\beta F = \lim_{N \to \infty} N^{-1} \ln Z$ in the thermodynamic
limit, where $N$ denotes the number of sites on the lattice.  For the uniform
case, we can write
\beq
Z = e^{((q/2)K + h)N} Z_r
\label{zrdef}
\eeq
where $q$ denotes the
coordination number of the lattice. For fixed $N$, $Z_r$ is then a polynomial
in $\mu$ and either $u$ (if $q$ is even) or $z$ (if $q$ is odd).
The 2D Ising model has the appealing feature of exact solvability;
the zero-field free energy $f$ and spontaneous magnetization $M$ were first
derived by Onsager and Yang, respectively \cite{ons,y} (both for the square
lattice; these solutions were later generalized to other 2D lattices).

   Yang and Lee pioneered a very interesting line of research in which one
studies the model with the external magnetic field generalized from real to
complex values \cite{yl,ly}.  For an arbitrary (finite as well as infinite)
lattice with ferromagnetic
spin-spin couplings $J_{n n'} \ge 0$ (and physical temperature $0 \le \beta \le
\infty$), Yang and Lee proved a classic theorem stating that the zeros of the
partition function lie on the unit circle $|\mu|=1$ in the $\mu$ plane and
pinch the positive real $\mu$ axis as the temperature $T$ decreases through
the critical temperature $T_c$ \cite{yl,ly} (see also
Ref. \cite{yrb}.) In the thermodynamic limit,
these zeros become dense and determine the continuous locus of points in the
$\mu$ plane where the free energy, for fixed $T$, is non-analytic.
Henceforth, we shall concentrate on the thermodynamic limit, and on results
 from finite graphs which can give insight into this limit.\footnote{
Several further questions
concerning these zeros have never been answered exactly.
These include, for the case of lattices
of dimensionality $d \ge 2$ (i) the density of zeros, and
(ii) the location of the complex-conjugate endpoints of the distribution for
$T > T_c$. Although we shall not need this here, we mention that there has
been intensive study of these zeros and generalizations since the original
works \cite{yl,ly}; some work (on regular lattices without quenched disorder)
is listed in Ref. \cite{further}.  For $J_{nn'} < 0$,
the zeros do not lie on the unit circle $|\mu|=1$ \cite{yw}. For the 1D
uniform--$J$ case, an exact solution by Yang and Lee \cite{ly,yw} shows that
these lie on the negative real $\mu$ axis; for $d \ge 2$, there are no exact
results on the location of the zeros for $J_{nn'} < 0$.
A numerical study was carried out in Ref. \cite{kay}.}  An
elementary property of (\ref{zfun}) is its invariance under the simultaneous
transformations $\sigma_n \to -\sigma_n$, $h \to -h$.  This implies that the
phase diagram in the $\mu$ plane is invariant under the inversion
\beq
\mu \to 1/\mu
\label{musym}
\eeq
with the obvious sign flip $M(\mu) = -M(1/\mu)$.
With no loss of generality, one may therefore restrict one's considerations
just to the interior and rim of the unit circle $|\mu|=1$ in the $\mu$ plane.

   A complementary complexification is to consider the model in zero field
with the temperature generalized to complex values.  In this case, one is
interested in the phase diagram in the complex $u$ plane (complex $z$ plane for
odd $q$).  A number of results in this area, both
exact and from series analyses, have been obtained for the Ising model with
spin 1/2 \cite{fisher}-\cite{cmo} (and with higher spin \cite{hse,egj,hs}).
The next logical step is to consider the model with both temperature
and magnetic field generalized to complex values.  Perhaps the earliest step
in this direction was the exact solution by Lee and Yang \cite{ly},
of the square-lattice Ising model for a particular manifold of pure imaginary
values of external field, $H = i(\pi/2)k_BT$, i.e. $h=i\pi/2$ ($\mu=-1$).
Recently, we have studied this area further \cite{ih,only}.

  In the present note we report some new results on complex-field (Yang-Lee)
zeros of the Ising model partition function.  We first prove a general
result. Consider the Ising model with uniform ferromagnetic spin couplings.
 For $d > d_{\ell.c.d.}=1$, the model has a low-temperature (i.e., small--$u$
for even $q$, small--$z$ for odd $q$) series expansion with a finite radius of
convergence. This fact allows one to carry out an analytic continuation from $u
\sim 0^+$ through $u=0$ to negative real values of $u$ in the vicinity of the
origin (for odd $q$, a continuation from $z \sim 0^+$ to negative real values
of $z$ near $z=0$).  The properties of the model are continuous
under this analytic continuation; in particular, the zeros of $Z$ remain on
the unit circle $|\mu|=1$.\footnote{We report results for real $u$ here.  We
have also calculated zeros for complex $u$; in this case, the zeros
in the $\mu$ plane are not symmetric under complex conjugation.  For example,
for the square lattice, we find zeros which lie along (i) a
spiral curve for $u=i$, and (ii) two linked, translated spiral-like curves for
$u=e^{i\pi/4}$.}
As we shall now show, the actual interval in negative $u$ or $z$
in which the zeros remain on $|\mu|=1$ depends on the lattice. We specialize to
(i) uniform nearest-neighbor couplings $J_{nn'}=J\delta_{n\pm e_i,n'}$ and (ii)
the thermodynamic limit (which, as usual, can be probed by using sufficiently
large finite lattices).

   We consider the square lattice first.  We found previously \cite{ih} that
for $\mu=-1$, the locus of points in complex temperature (the $u$ plane)
across which $f$ is non-analytic is the union of the unit circle and a finite
line-segment:
\beq
 \{ u= e^{i\theta} \} \cap  \{ 1/u_e \le u \le u_e \}
\label{sqline}
\eeq
where $0 \le \theta < 2\pi$ and
the inner endpoint, $u_e$, of the line segment is given by
\beq
u_e = -(3-2^{3/2}) = -0.171573...
\label{ue}
\eeq
Note that $u_e(\mu=-1) = -u_{c,sq}(\mu=1)$, where $u_{c,sq}$ is the critical
point separating the
$Z_2$--symmetric, paramagnetic (PM) phase and the broken-symmetry,
ferromagnetic (FM) phase of the $H=0$ model on the square lattice.
Now we switch our view back from the non-analyticities of $f$ in the
$u$ plane as a function of $\mu$ to the non-analyticities in the
$\mu$ plane as a function of $u$.  From our previous result (\ref{sqline}), we
know that as $u$ moves leftward from the origin along the negative real $u$
axis, there will be a point of non-analyticity when it reaches the
value $u_e$, and this point will occur, in the $\mu$ plane at $\mu=-1$.

   To
investigate the situation in the interval $u_e < u < 0$, we use two specific
methods: (i) exact calculation of the partition function and numerical
evaluation of the corresponding zeros in $\mu$ on finite lattices; and
(ii) analysis of low-temperature, high-field
series expansions.  All of our finite-lattice calculations for
$u_e < u < 0$ yield zeros in $\mu$ which, to within numerical accuracy
($\sim O(10^{-9}$) lie on the unit circle $|\mu|=1$.
In Fig. 1(a) we show a plot of zeros of $Z$ in $\mu$
for an $8 \times 8$ lattice with $u=-0.1$.  For this and the other figures,
periodic boundary conditions are used.  We have also made calculations with
helical and open boundary conditions, and obtain the same conclusions.

   As $u$ decreases through the value $u_e$, the situation becomes more
complicated.  We find that the zeros no longer all lie on the circle
$|\mu|=1$.  For $u$ slightly more negative than $u_e$, we find
that some zeros lie on the negative real $\mu$ axis (in a manner symmetric
under $\mu \to 1/\mu$, as implied by eq. (\ref{musym})).
This is illustrated by
Fig. 1(b), for $u=-0.25$.  In the thermodynamic
limit, these presumably merge to form a line segment which originates at the
point $\mu=-1$ when $u=u_e$ and spreads outward from this point as $u$ moves
to larger negative values.  Further structure may develop around $\mu=-1$ (see
Fig. 1(c) for $u=-0.5$). As $u$ becomes more negative, zeros appear
elsewhere, e.g. on the arcs evident in Fig. 1(c).

     We have confirmed the locations of these singularities by analyzing
low-temperature, high-field series expansions. With the definitions
\beq
f = (q/2)K + h + f_r(u,\mu)
\label{fdef}
\eeq
and $f_r = \lim_{N \to \infty} N^{-1} \ln Z_r$, these expansions can be
expressed as a small--$\mu$ series,
\beq
f_r = \sum_{n=1}^{\infty} L_n(u) \mu^n
\label{frmu}
\eeq
or a small--$u$ series,
\beq
f_r = \sum_{n=1}^{\infty} \psi_n(\mu) u^n
\label{fru}
\eeq
where $L$ and $\psi$ are polynomials in $u$ and
$\mu$, respectively.  For lattices with odd $q$, similar series hold with
$u$ replaced by $z$. Comparisons of series analyses with exact results,
e.g. for the 2D Ising model with $H=0$ ($\mu=1$) and $h=i\pi/2$ ($\mu=-1$; see
Ref. \cite{ih}) show that series of reasonable lengths yield very accurate
determinations of the positions of singular points.
The series (\ref{frmu}) has been computed to order
$\mu^{15}$ for the square (sq) lattice \cite{seg}
and to order $\mu^{12}$ for the triangular lattice \cite{seg,s75a}, to
be used below.  The series (\ref{fru}) has been calculated to $O(u^{23})$
for the square lattice \cite{be} and to $O(u^{21})$ for the triangular lattice
\cite{s73b}.  For a given value of $u$, we use the small-$\mu$ series to
compute the magnetization $M$ and fit this to a leading singularity
\beq
M_{sing} \sim (1-\mu/\mu_s)^{1/\delta_s}
\label{msingmu}
\eeq
where $\mu_s$ denotes a generic singular point (depending on $u$).  The
symmetry (\ref{musym}) automatically implies that $M$ then has the
singularity
\beq
M_{sing} \sim [(1-\mu/\mu_s)(1-\mu_s\mu)]^{1/\delta_s}
\label{msingmufull}
\eeq
As in our earlier papers \cite{chisq,ih,only}, we use dlog Pad\'e and
(first-order, unbiased) differential approximants for our study.  Details of
our methods are discussed in these papers.  We have checked our results by
analyzing the small-$u$ series (\ref{fru}), to compute $M$, fitting it
to a leading singularity of the form
\beq
M_{sing} \sim (1-u/u_s)^{\beta_s}
\label{msingu}
\eeq
where $u_s$ is a generic singular point (depending on $\mu$). That is, we
obtain a series of pairs of singular points $(\mu,u)_s$; these are singular
points in the $\mu$ plane for a given $u$ or, equivalently,
singular points in the $u$ plane for a given $\mu$.  The results
for the singular point $u_s(\mu)$ are in excellent agreement with the
values of $\mu_s(u)$ obtained from the small-$\mu$ series.
As we decrease $u$ from 0 through real values, we first find a firm
indication of a singularity when $u$ passes through $u_e$; this occurs at
$\mu=-1$.  This singularity moves inward toward the origin in the $\mu$ plane
as $u$ moves to the left of $u_e$. Some typical values are shown in Table I.

\begin{table}
\begin{center}
\begin{tabular}{|c|c|c|} \hline \hline & & \\
$u$ & $\mu_s$ & $1/\delta_s$ \\
& & \\ \hline \hline
$-1/4$ & $-0.615(10)$ & $-0.20(2)$ \\ \hline
$-1/3$ & $-0.400(5)$  & $-0.18(3)$ \\ \hline
$-1/2$ & $-0.219(2)$  & $-0.19(1)$ \\ \hline
$-2/3$ & $-0.1407(2)$ & $-0.20(1)$ \\ \hline
$-1$   & $-0.0735(1)$ & $-0.20(1)$ \\ \hline
$-3/2$ & $-0.0372(5)$ & $-0.20(1)$ \\ \hline
\hline
\end{tabular}
\end{center}
\caption{Values of $\mu_s$ and $1/\delta_s$ for various $u$ from our analysis
of the small--$\mu$ series for $M$ from (\ref{frmu}).  For $u=u_e$, the
location of the singular point is known exactly as $\mu_s=-1$.}
\label{table1 }
\end{table}

\begin{table}
\begin{center}
\begin{tabular}{|c|c|c|} \hline \hline & & \\
$\mu$ & $u_s$ & $\beta_s$ \\
& & \\ \hline \hline
$-0.15$ & $-0.6400(5)$ & $-0.185(15)$ \\ \hline
$-0.2$  & $-0.5314(5)$ & $-0.19(2)$   \\ \hline
$-0.4$  & $-0.3345(4)$ & $-0.19(2)$   \\ \hline
$-0.6$  & $-0.2529(3)$ & $-0.20(3)$   \\ \hline
$-0.8$  & $-0.2058(4)$ & $-0.21(2)$   \\ \hline
$-1$    & $u_e=-0.171573...$  & $-1/8$ \\ \hline
\hline
\end{tabular}
\end{center}
\caption{Values of $u_s$ and $\beta_s$ for various $\mu$ from our analysis
of the small--$u$ series for $M$ from (\ref{fru}).  The entries on the
last line are exact ($u_e$ is given in eq. (\ref{ue})).}
\label{table2 }
\end{table}

These results are in very good agreement with our findings from the
calculation of zeros of $Z$, and suggest that the singular points $\mu_s(u)$
and $1/\mu_s(u)$ are the inner
and outer endpoints of (what in the thermodynamic limit becomes a dense) line
segment of zeros of $Z$ (i.e., non-analyticities of $f$).

  As Table 1 shows, we
find that $M$ has a divergent singularity at $\mu_s$ (and hence, by
(\ref{musym}), also at $1/\mu_s$).  This is an analogue in the $\mu$ plane of
the general result in the $u$ plane which we have found in our previous work
on complex--$T$ singularities \cite{chisq,hs,ih} that $M$ diverges, as a
function of $u$, at the endpoints of
arcs or line segments of singularities protruding into the FM phase.  In
particular, for $\mu=-1$, the exact result \cite{ly} for $M(u)$
exhibits divergences at $u=u_e$ and $u=1/u_e$, the endpoints of the line
segment in (\ref{sqline}), with $\beta_e=-1/8$,
while from the exact result \cite{ly} for $f$ one can extract \cite{ih} the
specific heat exponent $\alpha_e'=1$.  Our series analysis \cite{ih} strongly
suggested the exact value $\gamma_e'=5/4$ at this point for the susceptibility
exponent (so that $\alpha_e'+2\beta_e+\gamma_e'=2$).  Although we have found
violations of scaling relations for complex--$T$ singularities, we note
that, provided the usual relations $\alpha'+\beta(\delta+1)=2$ and
$\gamma'=\beta(\delta-1)$ hold at $u=u_e$, it follows that
$1/\delta_e=-1/9$ at $(\mu,u)=(-1,u_e)$.
The fact that this differs from the value $1/\delta \sim
-0.2$ for the $(\mu,u)$ entries in Table 1 is not surprising, because
the point $\mu=-1$ is quite special, being related by a simple transformation
\cite{merlwu,ih} to the zero-field model, i.e. to the point $\mu=1$.  Thus,
just as a small nonzero value of the $H$ fundamentally changes the properties
of the zero field model ($H$ is a relevant parameter), i.e., the singular
properties change abruptly when $\mu$ moves slightly away from $\mu=1$,
so also, one expects
that a small change in $\mu$ away from $\mu=-1$ will have a similarly abrupt
effect on the singular properties.

   As the results in Table 2 show, $M$ is also divergent at the singular values
of $u_s$ corresponding to the $\mu$ values listed there.
The corresponding exponent $\beta \sim -0.2$, and, again, it is not
surprising that this differs from the exactly known value $\beta_e=-1/8$ at
$u=u_e$, $\mu=-1$, for the same reason as given above.

   For large negative $u$, we can observe that in the polynomial $L_n$ at
$n$'th order in $\mu$ in (\ref{frmu}), the dominant contribution is given
by the term of highest power in $u$, $\propto u^{q/2}$.  Now
consider the limit $|u| \to \infty$,  $\mu \to 0$ with
\beq
x \equiv u^{q/2} \mu
\label{xvar}
\eeq
fixed.  Then the double series (\ref{frmu}) for $f_r$ reduces to a series in
the single variable $x$.  We have analyzed this series to determine the
singular point $x_s$.  For a given large negative value of $u$, we can then
extract the asymptotic value of the singularity $\mu_s$.  Clearly, as $u \to
-\infty$, $\mu_s$ approaches the origin like
\beq
\mu_s \sim a u^{-q/2}
\label{musasymp}
\eeq
where $a$ is a constant (and $q=4$ and 6 for the sq and tri lattices).

   We also observe from our calculations of zeros that for small negative $u$
(e.g. Fig. 1), the density $g(\theta)$ of zeros is consistent with being
constant on the circle.
Although $g(\theta,u)$ is not known exactly even in the
Yang-Lee region $0 \le u \le 1$ or the subinterval of this region lying in the
FM phase for the square lattice, viz., , $0 \le u \le u_c$ (in which the zeros
cover the entire circle), we have found by explicit calculation on finite
lattices that for $u$ in this latter interval that as $u \to 0^+$, the
density $g(\theta)$ is again consistent with being constant.

   We have also calculated zeros for real $u < 1$. These
fall in more complicated patterns.
Here our emphasis is on determining the boundaries of the intervals
in negative $u$ where the zeros still lie on the unit circle $|\mu|=1$ and
establishing the nature of the singularities (which we show are line segments
on the real $\mu$ axis) which first appear when these zeros start moving off
the circle.

   Next, we carry out an analogous study for the triangular lattice
of the size of the interval along the negative $u$ axis for
which the zeros stay exactly on the circle $|\mu|=1$.
We bring to bear our knowledge of the complex-temperature phase
diagram: for $h=0$ ($\mu=1$), the continuous locus of points where $f$ is
non-analytic is the union of the circle and semi-infinite line \cite{chisq}
\beq
\{ u=-1/3 + (2/3)e^{i\theta} \} \cap \{-\infty \le u \le -1/3 \}
\label{utri}
\eeq
where $0 \le \theta < 2\pi$.
Recall that the physical critical point is $u_{c,t}=1/3$.  For $\mu=-1$, the
non-analytic points in the $u$ plane consist of a circular arc and
semi-infinite line segment \cite{only}
\beq
\{u = (1/2)(-1 + e^{i\theta}) \ ; \ \theta_{ce} \le |\theta| \le \pi \} \cap
\{ -\infty \le u \le -1/2 \}
\label{utrimum1}
\eeq
where $u_{ce}=e^{i\theta_{ce}}=(1/9)(-1 + 2^{3/2}i)$. From explicit
calculations of zeros in $\mu$ and series analysis, we find that for
$-1/3 \le u < 0$, the zeros of $Z$ continue to lie on the circle $|\mu|=1$.
(The positions are exactly on this circle to within the numerical accuracy of
$\sim O(10^{-9})$.)  Fig. 2(a) is a plot of zeros for
$u=-0.1$ on a triangular lattice\footnote{The triangular lattice
can be represented by a square lattice with an interaction between, e.g., the
spins on the lower-left and upper-right corners of each square.  We use a $6
\times 6$ lattice of this type with periodic boundary conditions.}.
As $u$ decreases through $-1/3$ moving to larger negative
values, some zeros appear on the positive real $\mu$ axis, starting at $\mu=1$
and spreading outward along the positive real axis from this point.  Fig. 2(b)
shows the zeros for $u=-0.4$.  It is plausible that in the thermodynamic
limit, these
merge to form line segments. We have used the small-$\mu$ and small-$u$
series to determine the position of the inner endpoint, $\mu_{e,rhs}$, of
this line segment (and hence, by the $\mu \to 1/\mu$ symmetry, also the
outer endpoint).  We find results in agreement with the zeros calculated on the
finite lattice.  Here, $lhs$, $rhs$ denote ``left-, right-hand side''.
As $u$
reaches $-1/2$, our exact results in Ref. \cite{only} show that there is a new
set of singularities in the $\mu$ plane first appearing at $\mu=-1$.  This
singularity $\mu_{e,lhs}$, moves inward toward the origin along the negative
real $\mu$ axis (and its inverse moves outward) as $u$ decreases past $-1/2$
toward larger negative values.  We find that the small-$\mu$ series
(\ref{frmu}), to the order calculated, is not sensitive to $\mu_{e,rhs}$;
however, the small-$u$ series (\ref{fru}) does allow a rough determination of
its value.

    Finally, we prove a theorem concerning the reduced free energy for the
honeycomb (hexagonal) lattice and for two heteropolygonal lattices: the
$3 \cdot 12^2$ and $4 \cdot 8^2$ lattices.  (For notation, we refer the reader
to our previous paper on heteropolygonal lattices \cite{cmo}.)  These
lattices all have odd coordination number ($q=3$).
\vspace{2mm}

Theorem: For lattices with odd coordination number $q$,
\beq
Z_r(z,-\mu) = Z_r(-z,\mu)
\label{frel}
\eeq
This is proved as follows.
Now $\mu \to -\mu$ corresponds to $h \to h + i\pi/2$.  Using the identity
$e^{i(\pi/2)\sigma_n} =i\sigma_n$, we have
\beq
Z(K,h+i\pi/2) = \sum_{\{\sigma\}}\biggl ( \prod_{<nn'>}e^{K\sigma_n\sigma_{n'}}
\biggr ) \biggl ( \prod_n i \sigma_n e^{h\sigma_n} \biggr )
\label{zshift}
\eeq
Since $q$ is odd, we may replace $\sigma_n = \sigma_n^q$ on each site.  We can
then associate each factor in $\sigma_n^q$ with one of the $q$ bonds adjacent
to this site. The product of $\sigma_n^q$
over sites is thus re-expressed as a product of pairs
$\sigma_n \sigma_{n'}$ over bonds.  We next use the identity
$e^{i(\pi/2)\sigma_n\sigma_{n'}} = i\sigma_n \sigma_{n'}$ to re-express
this product over bonds in terms of an exponential with shifted coupling
$\tilde K$:
\beq
Z(K,h+i\pi/2) = i^{(2-q)N/2)} \sum_{\{\sigma\}}
e^{\tilde K\sum_{nn'}\sigma_n\sigma_{n'} + h\sum_n\sigma_n } =
i^{(2-q)N/2} Z(\tilde K, h)
\label{zkt}
\eeq
where
\beq
\tilde K = K + \frac{i\pi}{2}
\label{kshift}
\eeq
The shift (\ref{kshift}) takes $z=e^{-2K}$ to $-z$; using this together
with (\ref{zrdef}) then yields the theorem.

   A consequence of this theorem is that for a lattice with odd $q$, the zeros
of the partition function in the $\mu$ plane lie on the unit circle $|\mu|=1$
not just for the Yang-Lee range $0 \le z \le 1$, but for the larger range
\beq
-1 \le z \le 1
\label{fullrange}
\eeq
In the thermodynamic limit, for $z_c < z \le 1$ or $-1 \le z < -z_c$,
(where $z_c = e^{-2K_c}$ is the respective critical point in the $z$ variable
for each such lattice)
these zeros form a circular arc which does not completely enclose the origin in
the $\mu$ plane, allowing an analytic contination from the interior of the unit
circle to its exterior.  As $z$ decreases through $z_c$ or $-z$ increases
through $-z_c$ along the real $z$ axis,
the arc closes to form a complete circle which prevents an
analytic continuation from the interior to the exterior of the unit circle
$|\mu|=1$.  In passing, we recall that for the honeycomb lattice, $z_c =
2-\sqrt3$ (see Ref. \cite{cmo} for the complex-temperature phase diagrams of
the $z_c$ values for the $3 \cdot 12^2$ and $4 \cdot 8^2$ lattices).

   This research was supported in part by the NSF grant PHY-93-09888.

\begin{center}
{\bf Figure Captions}
\end{center}

 Fig. 1. \ Zeros of $Z$ in the $\mu$ plane for the Ising model on a square
lattice of size $8 \times 8$ for $u=$ (a) $-0.1$, (b) $-0.25$, (c) $-0.5$.

 Fig. 2. \ Zeros of $Z$ in the $\mu$ plane for the Ising model on a triangular
lattice of size $6 \times 6$ for (a) $u=-0.1$, (b) $u=-0.4$.


\begin{thebibliography}{99}

\bibitem{ons}{L. Onsager, Phys. Rev. {\bf 65} (1944) 117.}

\bibitem{y}{C. N. Yang, Phys. Rev. {\bf 85} (1952) 808.}

\bibitem{yl}{C. N. Yang and T. D. Lee, Phys. Rev. {\bf 87} (1952) 404.}

\bibitem{ly}{T. D. Lee and C. N. Yang, Phys. Rev. {\bf 87} (1952) 410.}

\bibitem{yrb}{C. N. Yang, {\it Selected Papers 1945-1980 with Commentary}
(Freeman, 1983), pp. 14-16.}

\bibitem{yw}{C. N. Yang, {\it Special Problems of Statistical Mechanics}
(Lectures at the University of Washington, 1952), pp. 172-173.  We thank
Professor Yang for kindly lending a copy of these unpublished lectures to us.}

\bibitem{further}{M. Suzuki, Prog. Theor. Phys. {\bf 38} (1967) 1225; {\bf 40}
(1968) 1246; {\bf 41} (1969) 1438; T. Asano, {\it ibid.} {\bf 40} (1968) 1328;
Phys. Rev. Lett. {\bf 24} (1970) 1409;
R. B. Griffiths, J. Math. Phys. {\bf 10} (1969) 1559 ;
M. Suzuki, C. Kawabata, S. Ono, Y. Karaki, and M. Ikeda,
J. Phys. Soc. Jpn. {\bf 29} (1970) 837;
M. Suzuki and M. E. Fisher, J. Math. Phys. {\bf 12} (1971) 235;
P. Kortman and R. B. Griffiths, Phys. Rev. Lett. {\bf 27} (1971) 1439;
S. Katsura and M. Ohminami, J. Phys. A {\bf 5} (1972) 95;
G. A. Baker, Jr. and J. Moussa, J. Appl. Phys. {\bf 49} (1978) 1360;
M. E. Fisher, Phys. Rev. Lett. {\bf 40} (1978) 1610;
D. Kurtze and M. E. Fisher, J. Stat. Phys. {\bf 19} (1978) 203;
G. A. Baker, Jr., M. E. Fisher, and P. Moussa, Phys. Rev. Lett. {\bf 42}
(1979) 615; E. H. Lieb and A. Sokal, Commun. Math. Phys. {\bf 80} (1981) 153;
J. Cardy, Phys. Rev. Lett. {\bf 54} (1985) 1354.}

\bibitem{kay}{S. Katsura, Y. Abe, and M. Yamamoto, J. Phys. Soc. Jpn.
{\bf 30} (1971) 347.}

\bibitem{fisher}{M. E. Fisher, {\it Lectures in Theoretical Physics}
(Univ. of Colorado Press), vol. 7C (1965), p. 1.}

\bibitem{ct}{S. Katsura, Prog. Theor. Phys. {\bf 38} (1967) 1415;
S. Ono, Y. Karaki, M. Suzuki, and C. Kawabata
J. Phys. Soc. Jpn. {\bf 25} (1968) 54;
A. J. Guttmann, J. Phys. C {\bf 2} (1969) 1900;
C. Domb and A. J. Guttmann, J. Phys. C {\bf 3} (1970) 1652;
C. Itzykson, R. Pearson, and J. B. Zuber, Nucl. Phys. {\bf B220} (1983)
415; G. Marchesini and R. Shrock, Nucl. Phys. {\bf B318} (1989) 541.}

\bibitem{egj}{I. G. Enting, A. J. Guttmann, and I. Jensen, J. Phys. A
{\bf 27} (1994) 6987.}

\bibitem{chisq}{V. Matveev and R. Shrock, J. Phys. A {\bf 28} (1995) 1557;
{\it ibid.}, in press.}

\bibitem{cmo}{V. Matveev and R. Shrock, J. Phys. A {\bf 28} (1995) 5235.}

\bibitem{hse}{C. Kawabata and M. Suzuki, J. Phys. Soc. Jpn. {\bf 27} 1105
(1969); P. F. Fox and A. J. Guttmann, J. Phys. C {\bf 6} (1973) 913.}

\bibitem{hs}{V. Matveev and R. Shrock, J. Phys. A. (Lett.) {\bf 28} (1995)
L533;  Phys. Lett. {\bf A204} (1995) 353.}

\bibitem{ih}{V. Matveev and R. Shrock, J. Phys. A {\bf 28} (1995) 4859.}

\bibitem{only}{V. Matveev and R. Shrock, Phys. Rev. E, in press
(cond-mat/9507120).}

\bibitem{seg}{M. F. Sykes, J. W. Essam, and D. S. Gaunt, J. Math. Phys.
{\bf 6} (1965) 283;
M. F. Sykes, D. S. Gaunt, S. R. Mattingly, J. W. Essam, and
C. J. Elliott, J. Math. Phys. {\bf 14} (1973) 1066.}

\bibitem{s75a}{M. F. Sykes, S. McKenzie, M. G. Watts, and D. S. Gaunt,
J. Phys. A {\bf 8} (1975) 1461.}

\bibitem{be}{R. J. Baxter and I. G. Enting, J. Stat. Phys. {\bf 21}
(1979) 103.}

\bibitem{s73b}{M. F. Sykes, D. S. Gaunt, J. L. Martin, S. R. Mattingly, and
J. W. Essam, J. Math. Phys. {\bf 14} (1973) 1071;
M. F. Sykes, M. G. Watts, and D. S. Gaunt, J. Phys. A {\bf 8} (1975) 1448.}

\bibitem{merlwu}{D. Merlini, Lett. Nuovo Cim. {\bf 9} (1974) 100;
K. Y. Lin and F. Y. Wu, Int. J. Mod. Phys.  {\bf B4} (1988) 471.}

\end{thebibliography}
\end{document}